# Giant mid-IR resonant coupling to molecular vibrations in sub-nm gaps of plasmonic multilayer metafilms

Rakesh Arul, David Benjamin-Grys, Rohit Chikkaraddy, Niclas S Mueller, Angelos Xomalis, Ermanno Miele, Tijmen G Euser, and Jeremy J Baumberg[*]

[1] NanoPhotonics Centre, Cavendish Laboratory, Department of Physics, JJ Thompson Avenue, University of Cambridge, Cambridge, CB3 0HE, United Kingdom

[*] email: jjb21@cam.ac.uk

**Abstract**

**Nanomaterials capable of confining light are desirable for enhancing spectroscopies such as Raman scattering, infrared absorption, and nonlinear optical processes. Plasmonic superlattices have shown the ability to host collective resonances in the mid-infrared, but require stringent fabrication processes to create well-ordered structures. Here, we demonstrate how short-range-ordered Au nanoparticle multilayers on a mirror, self-assembled by a sub-nm molecular spacer, support collective plasmon-polariton resonances in the visible and infrared, continuously tunable beyond 11 µm by simply varying the nanoparticle size and number of layers. The resulting molecule-plasmon system approaches vibrational strong coupling, and displays giant Fano dip strengths, SEIRA enhancement factors ∼$10^6$, light-matter coupling strengths $g$∼100 cm$^{-1}$, Purcell factors ∼$10^6$, and mode volume compression factors ∼$10^8$. The collective plasmon-polariton mode is highly robust to nanoparticle vacancy disorder and is sustained by the consistent gap size defined by the molecular spacer. Structural disorder efficiently couples light into the gaps between the multilayers and mirror, enabling Raman and infrared sensing of sub-picolitre sample volumes.**

**Introduction**

Infrared (IR) and Raman spectroscopies provide complementary probes of molecular vibrations, and are widely used to sense molecules. However, only one in a million photons are inelastically scattered by the Raman process, and the interaction of vibrational transitions with IR light is an order-of-magnitude weaker than with electronic transitions in the visible region[1]. This limits the ability of conventional IR and Raman spectroscopies to measure trace analytes and demonstrates the need for an effective way to strengthen light-matter interactions in order to measure picomolar concentrations. Plasmonic constructs enhance the interaction of light with molecular vibrations[2] using micron-scale resonant antennae[3], metamaterials[4], and nanoparticle aggregates[5], and have been used to detect molecular structure and conformation via surface-enhanced IR absorption (SEIRA) spectroscopy[6] for various chemical, biological, and security applications. Infrared confinement is also useful in light-harvesting for photovoltaics[7] or photocatalysis, thermal management metasurfaces[8, 9], and in the design of new infrared sources and detectors[10]. Metal plasmonic nanocavities are highly efficient in reducing the mode volume of light, operate across a broad spectrum from the UV to THz[11-13], and can be fabricated via self-assembly[14-19] unlike surface phonon-polaritons[20] and acoustic graphene plasmons[21]. However, for wavelengths $\lambda$ in the mid-infrared (MIR), plasmonic cavity mode volumes achieved so far are on the order of $(\lambda)^3$ in microcavities[3] or $100 \cdot (\lambda)^2$ in patch antennae[22]/superlattices[19]. Plasmonic nanoparticle resonances can be tuned towards the infrared by assembling nanoparticle chains[11], which creates a collective super-radiant mode that saturates at $\lambda \sim 1$ μm[12] due to their intrinsic coupling. By extending the lateral extent [14-17] and number of layers to create well-ordered superlattices[19] or superparticles[13], these resonances can be pushed beyond 1 μm. However, highly ordered nanoparticle superlattices are difficult to fabricate with small gap sizes, and the tortuosity of their gaps and interstices makes bringing analyte molecules into optical hotspots difficult. Randomly deposited gold nanoparticle aggregates[23] and rough gold substrates[24] possess broad enhancements in the infrared spectral window for SEIRA, and lack specific resonances. The need to develop better SEIRA-active substrates has been identified[2] (since developments lag behind SERS substrates), as well as to develop combined SERS-SEIRA platforms.

Here we show giant (>90% extinction) mid-infrared plasmonic resonances which have much smaller mode volumes of $10^{-8}(\lambda)^3$ and are highly tuneable. These are created by hierarchical self-assembly of amorphous gold nanoparticle multilayer films on a mirror (NP$n$ML-on-mirror). Each layer is a disordered two-dimensional nanoparticle aggregate made by linking gold nanoparticles (AuNPs) with the molecular glue cucurbit[5]uril (CB[5])[25], with the layers then sequentially deposited onto a gold mirror. These multilayer NP$n$ML-on-mirror films, with $n$ indicating the number of layers, possess dual resonances in the visible and infrared regions. Their dominant infrared resonance can be tuned beyond $\lambda$>11 μm by increasing the AuNP size and number of layers, and is remarkably robust to disorder. Furthermore, free-space coupling of light into these nanogaps is efficient, making them practical for realistic SEIRA applications. The amorphous AuNP arrangement with precise nanogap sizes[25] localizes visible and infrared light to sub-wavelength volumes, improves in/out-coupling of light, and allows giant SEIRA (>$10^6$) and SERS ($10^6$) enhancements to be obtained greatly exceeding all existing self-assembled platforms (benchmarked in Supplementary Tables S1 and S2).

## Results and Discussion

### Tuning collective plasmonic modes to the mid-infrared

Millimetre-wide monolayers of nanoparticles with diameter $D$ = 20-100 nm are fabricated by drop-casting a concentrated aqueous solution of AuNPs that are aggregated with CB[5] and concentrated in chloroform (see Methods). Since aggregation occurs at the chloroform-water interface, a two-dimensional AuNP sheet is produced with fill fraction ~65%. The resulting films are characterized to determine their surface morphology (Fig. 1a-c), showing separate regions with 1 or 2 layers. These dense layers are built from disordered AuNP networks (Supplementary Information Section S1) forming connected clusters with many vacancies. Cucurbituril-binding is crucial to achieve $d$=0.9±0.05 nm gap sizes[25] that result in prominent resonances deep in the MIR. The gap size inhomogeneity when instead using salt or acid aggregation results in much broader modes (Supplementary Information Section S2). By depositing NP droplets sequentially after each previous one dries, NP$n$ML multilayers ($n$=1-9) can be rapidly constructed with a layer-number-dependent optical response (Fig.1d,e).

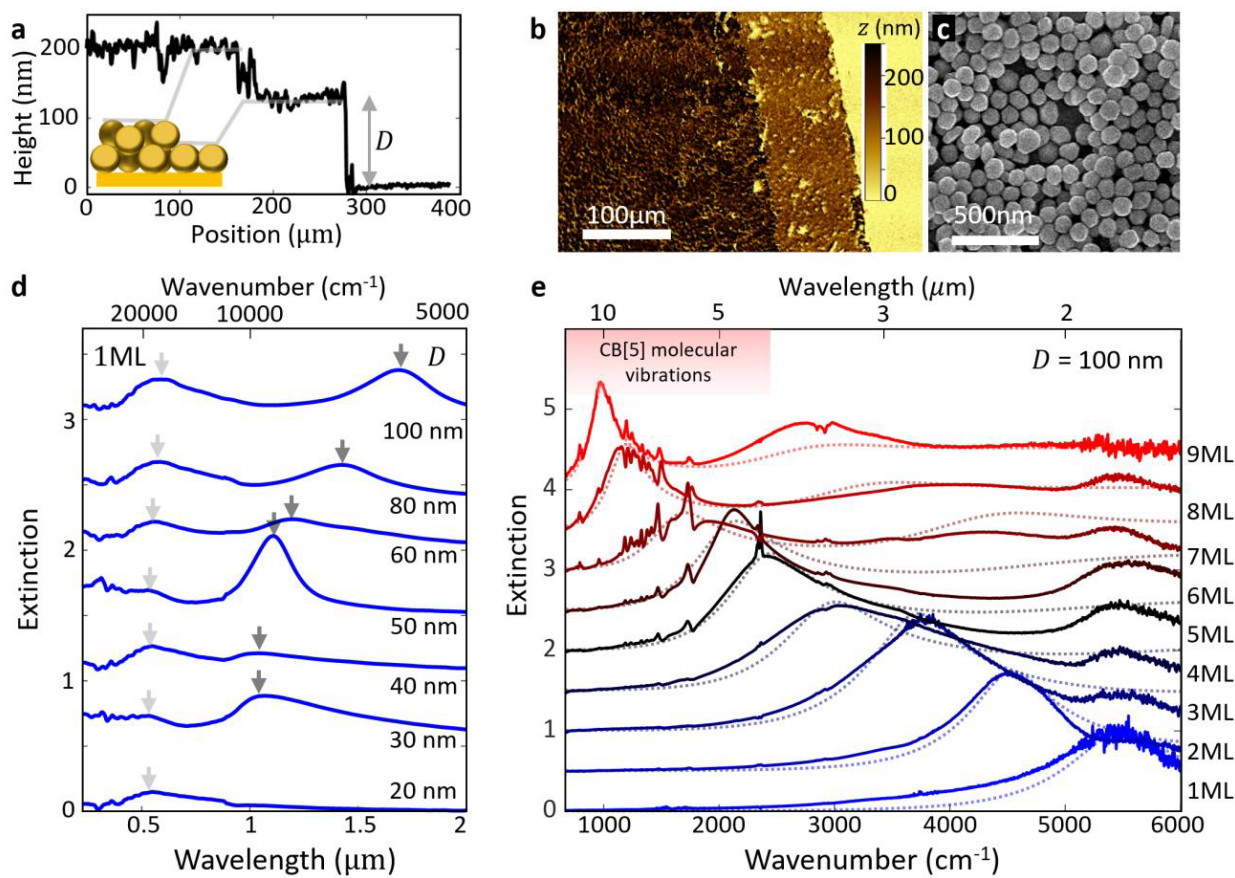

**Figure 1. Optical response of amorphous AuNP multilayers on gold (NP$n$ML-on-mirror).** (a) Height profile of monolayer (1ML) and bilayer (2ML) disordered AuNP ($D$=100nm) multilayer films from optical profilometer. (b) Optical profilometer image of monolayer and bilayer films. (c) Scanning electron micrograph of 2ML film. (d) Extinction spectra of 1ML films with increasing AuNP diameter $D$ = 20 to 100 nm. Arrows mark resonance positions. (e) FTIR extinction spectrum of NP$n$ML-on-mirror films for $n$ = 1 to 9 with AuNP diameter $D$ = 100 nm. Red shaded represents spectral region with CB[5] vibrational

absorption peaks. All spectra expressed in absorbance units ($-\log_{10} R$). Dotted lines show fits from collective plasmon-polariton model (SI Section S3).

The optical properties of the NP1ML-on-mirror monolayer films vs NP diameter $D$ are measured in a custom-built visible (VIS) to near-infrared (NIR) reflectance spectrometer (Fig. 1d). Two plasmonic modes are found in the VIS and NIR, both redshifting with increasing NP size. A AuNP diameter of 100 nm is sufficient to tune the single layer resonance to $\lambda$=1.7 µm. By then increasing the film thickness using $n$ from 1 to 9 layers, the resonance tunes further into the MIR, out to $\lambda$=11 µm (27 THz, 900 cm$^{-1}$ in Fig. 1e). These resonances also display near complete extinction, with reflectance < 5%. Simultaneously, the infrared absorption bands of the CB[5] molecules embedded in the nanogaps are enhanced significantly when the MIR resonance passes through their vibrational transitions (Fig. 1e).

**Origin of mid-infrared mode tuning**

Individual nanoparticle plasmons within well-ordered face-centered cubic AuNP superlattices couple with each other to form a collective plasmonic mode $\widetilde{\omega}_p$ [18, 26]. A simple estimate of the 1ML plasmon frequency (Fig.1d) from a generalized circuit model in the small gap limit[27] indicates that $\widetilde{\omega}_p$ is set by the capacitance of the AuNP nanogaps (Supplementary Information Section S3), but does not change significantly with gap size. This collective plasmon $\widetilde{\omega}_p$ then hybridizes with propagating light within the film $\omega_l(k_n) = ck_n/n_{\text{eff}}$ (where $n_{\text{eff}}$ is the effective dielectric constant in this material), forming a bulk plasmon-polariton mode ($\omega_{\text{BPP}}$) with dispersion set by the light-plasmon interaction strength ($\Omega$)[18, 26]:

$$2\omega_{\text{BPP}}^2 = \omega_l^2(k) + \widetilde{\omega}_p^2 + 4\Omega^2 - \sqrt{\left[\omega_l^2(k) + \widetilde{\omega}_p^2 + 4\Omega^2\right]^2 - 4\widetilde{\omega}_p^2 \omega_l^2(k)} \tag{1}$$

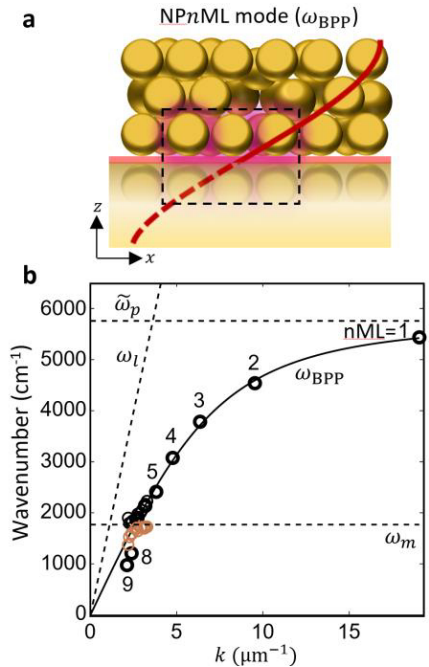

**Figure 2. Origin of infrared modes and field localization.** (a) Mode from layers of AuNP lattice-on-mirror (generating image dipoles). Red line depicts compression of IR $\lambda$ from free space within the NP$n$ML-on-mirror, box indicates septamer-on-mirror motif. (b) Dispersion of NP$n$ML-on-mirror plasmon polariton $\omega_{\text{BPP}}$ (points - experimental data for $n$ = 1 to 9 MLs) and fit to Hopfield model (black line). Black and brown circles show additional anticrossing (see Fig. 3) from coupling to CB[5] C=O stretch ($\omega_m$). Dashed lines show light-line inside uncoupled NP$n$ML-on-mirror ($\omega_l$) and bulk collective plasmon frequency ($\widetilde{\omega}_p$).

In a multilayer of finite thickness, plasmon-polaritons reflect off the bottom and top surfaces to form standing waves, giving rise to pronounced Fabry-Perot-like optical resonances[18, 19]. The giant absorption of NP$n$ML-on-mirror films (reflection < 5%), relative to the perfectly ordered case (Supplementary Information Fig. S8), is due to the bottom mirror preventing transmission and the disorder-induced scattering loss[28]. The mirror at the bottom surface enforces a field-null here, which sets the fundamental Fabry-Perot mode resonant condition $\lambda_n = 4L$, where $L = \sqrt{\frac{2}{3}} nD$ is the thickness of the NP$n$ML[18]. This NP$n$ML-on-a-mirror metamaterial slab confines a quarter-wavelength within it (Fig. 2a), resulting in wave-vector $k_n = 2\pi/(4L)$ of the collective plasmonic mode (Fig. 2b).

Using this condition, the observed plasmon resonances for $n$=1-9ML (Fig. 1e) compare well with the dispersion model of equation (1) (Fig. 2b), explaining the red shift with increasing nanoparticle diameter and layer number (Fig. 1d,e). A good fit is obtained (Fig. 2b black line) with $n_{\text{eff}}$=1.3, $\widetilde{\omega}_p$=0.71 eV, and $\Omega$=0.51 eV (similar to that in highly ordered AuNP superlattices[18]). When $\omega_{\text{BPP}}$ becomes resonant with the CB[5] C=O stretch at $\omega_m$ = 1765 cm$^{-1}$ (Fig. 2b,3b), an additional anti-crossing/coupling of $\omega_{\text{BPP}}$ with $\omega_m$ occurs. As we will show below, the MIR light trapped in these modes occupies extremely small mode volumes, giving strong SEIRA spectra near resonance (Fig.1e).

**Coupling of molecular vibration to collective plasmon**

For NP7ML-on-mirror films, the broad plasmonic resonance (~2100 cm$^{-1}$) results in a giant enhancement of the nearby $\omega_m$ = 1765 cm$^{-1}$ carbonyl (C=O) barrel-stretch mode of CB[5] (Supplementary Information Section S5). The IR absorption bands display a highly asymmetric Fano lineshape due to coupling and interference with the plasmonic mode[29]. From X-ray photoelectron spectroscopy measurements of these films (Supplementary Information Section S6), the upper limit of CB[5] coverage is 0.89nm$^{-2}$ (inter-molecular spacing = 1.14 nm). Comparing to infrared transmission spectra of 5mM CB[5] in water gives a lower bound of the SEIRA enhancement factor (EF) as (1.2±0.1)x10$^6$ (Supplementary Information Section S7), performing better than state-of-the-art SEIRA platforms[3, 30-32] which only reach enhancement factors between 10$^2$-10$^4$ (detailed comparison in SI Tables S1 and S2).

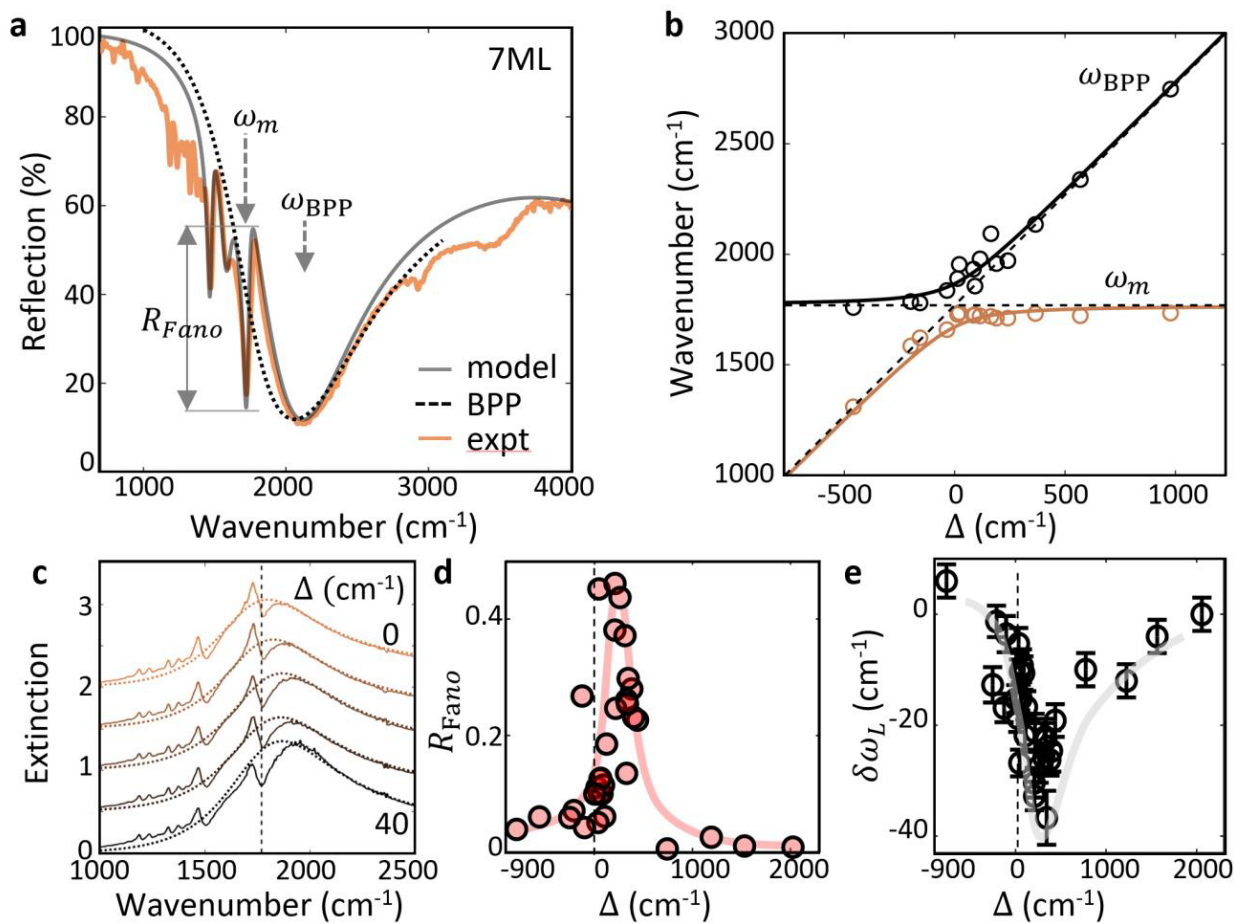

**Figure 3. Vibrational-plasmon coupling.** (a) Fano lineshape of CB[5] vibration in coupled oscillator model fit ('model', black line, SI Section S3) to the measured reflectance spectrum ('expt', blue line) of NP7ML-on-mirror films. (b) Extracted coupled plasmon-vibrational modes show an anticrossing with coupling strength $g \sim 100$ cm$^{-1}$. (c) Extinction for detunings around resonance, showing phase change of Fano dip. (d) Fano dip strength $R_{Fano}$ of the SEIRA molecular absorption at 1765 cm$^{-1}$, and (e) molecular Lamb shift $\delta\omega_L$ vs plasmon-molecular vibration detuning $\Delta = \omega_{BPP} - \omega_m$ between molecular ($\omega_m$) and plasmon ($\omega_{BPP}$) resonances (lines are guides to eye).

By extracting the location of the molecular Fano dip and the centre of the broad plasmonic resonance (Supplementary Information Fig. S18), an anticrossing can be seen as the plasmonic resonance tunes across the 1765 cm$^{-1}$ vibration (Supplementary Information Fig. S19a, with changing film thickness around 7-8 ML). The optomechanical (vibration-plasmon) coupling strength $g = 102 \pm 8$ cm$^{-1}$ (Supplementary Information Fig. S19b) is comparable to many realizations of carbonyl bond (C=O) vibrational strong coupling with previous $g$ ranging from 30[33], 54[22], 64[34], to 75[35] cm$^{-1}$. However, due to the large linewidth of the plasmon ($\gamma_c \sim 770$ cm$^{-1}$) and small molecular linewidth ($\gamma_m \sim 70$ cm$^{-1}$), the coupled system remains in the weak coupling[36] regime, defined by $2g/\gamma_c \sim 0.3$ (thus < 1). On the other hand, the effective cooperativity factor $C = 4g^2/\gamma_c\gamma_m \sim 0.8$ is comparable to single-molecule strong coupling in the visible[37] and mid-infrared collective strong coupling[33].

The Fano lineshape can be fitted by a modified coupled oscillator model[38], which accounts for the change in refractive index of the plasmonic system due to coupling with molecular resonances in the local Green's function (Supplementary Information Section S8) as a modified polarization,

$$P(\omega) \propto \frac{(\nu+q)^2 + B}{(\nu^2 + 1)} \quad (2)$$

with normalized detuning $\nu = [\omega^2 - (\omega_m + \delta\omega_L)^2]/\omega\gamma_m$, Fano asymmetry parameter $q = 2(\delta\omega_L - \delta\omega_L')/\gamma_m$ and $B = (\gamma_m'/\gamma_m)^2$, where $\omega$ is the incident frequency, $\omega_m$ the molecular transition frequency, $\delta\omega_L$ the molecular Lamb shift, $\gamma_m$ the molecular decay rate, and $\delta\omega_L'$, $\gamma_m'$ the multipolar plasmonic modifications to the Lamb shift and decay rates respectively (see [43] and Supplementary Information Section S8). Fitting Eqn. (2) allows coupling parameters to be extracted as a function of detuning (Fig.3d,e).

The strength of the plasmon-molecule interaction is observed in the magnitude of the Fano peak-to-dip extinction ($R_{\text{Fano}}$), which can be almost as large as the plasmon mode extinction (Fig. 3a,c,d), and the molecular Lamb shift ($\delta\omega_m$), which can reach up to 40 cm$^{-1}$. This greatly exceeds any existing SEIRA platforms which have $R_{\text{Fano}}$ of only a few percent[3, 30-32] (SI Table S1 and S2). Furthermore, there is a strong multipolar contribution to the near-field Green's function, beyond the dipolar approximation[38], which introduces an additional Lamb shift ($\delta\omega_L'$) and linewidth broadening ($\gamma_m'$). This results in an asymmetric spectrum at zero-detuning $\Delta = \omega_{\text{BPP}} - \omega_m \sim 0$ (Fig. 3c), where a symmetric one is normally expected. Hence, the largest $\delta\omega_m$ and $R_{\text{Fano}}$ do not occur at zero detuning, but at $\Delta \sim 200$ cm$^{-1}$. Finally, the Fano asymmetry parameter ($q$) of the SEIRA peaks flip sign[39] as expected, depending on the detuning $\Delta$ (Supplementary Figure S19c).

From the coupling strength and radiative decay rates (Supplementary Information Section S9), the effective radiative Purcell enhancement factor $F_p = (\gamma_m - \gamma_m')/\gamma_m^S = 4g^2/\gamma_c\gamma_m^S$ can be found[38], where the spontaneous radiative decay rate of CB[5] outside the cavity is $\gamma_m^S$. This spontaneous molecular radiative decay can be extracted from the oscillator strength of the CB[5] vibration ($\gamma_m^S$ = 1.1x10$^{-6}$ cm$^{-1}$, Supplementary Information Section S9), and the decay within the cavity from fitting the spectrum in Figure 3a ($\gamma_m$ = 62±5 cm$^{-1}$, $\gamma_m'$ = 49±4 cm$^{-1}$). The resulting average Purcell factor $F_P$ = 6±1 x10$^6$ is large (even exceeding 10$^7$) and comparable to many single nanostructure Purcell factors in the visible regime[40]. The mode volume ($V_m$) reduction compared to free-space corresponding to this Purcell factor is $V_m$ = 1.5x10$^{-8}$ $\lambda^3$ ($V_m$ = 2900 nm$^3$). The extracted mode volume thus implies light confinement within ~7 gaps assuming typical 25 nm diameter facets[27] and $d$=0.9 nm gap widths defined by CB[5]. This is consistent with the number of gaps within a AuNP septamer, which forms the typical structural motif in the disordered film (Fig. 1b and Supplementary Information Fig. S1), and approaches the limit of tightest confinement possible within this nanostructure. Squeezing the near-field into consistent gaps of spacing 0.9 nm accounts for the large optomechanical coupling strength $g$, the dip strength $R_{\text{norm}}$, Lamb shift $\delta\omega_m$, and the giant SEIRA enhancement factors measured.

## SEIRA – SERS comparison on a single substrate

In the following, we compare the SEIRA and SERS spectra from a NP7ML-on-mirror film. While the SEIRA spectrum is dominated by the portal C=O stretch of CB[5] in the gaps (Fig. 4a), the SERS-active vibrations have different selection rules and instead are dominated by the expected 830 cm$^{-1}$ peak from the CB[5] ring-breathing mode[41] (Fig. 4b). In addition, multiple peaks from the citrate capping agent used to stabilize AuNPs are observed only in the SERS spectrum. Hyperspectral IR and Raman maps of the same substrate show substantially different enhancements in the bands associated with CB[5]. The SEIRA C=O map (Fig. 4c) has different local intensities due to the heterogeneity in number of AuNP layers which locally tunes the $\omega_{\text{BPP}}$ resonance. The SERS map (Fig. 4d) is more homogeneous because the SERS enhancement depends more on the local septamer structure of the AuNPs in the visible regime and the homogeneous nanogap sizes.

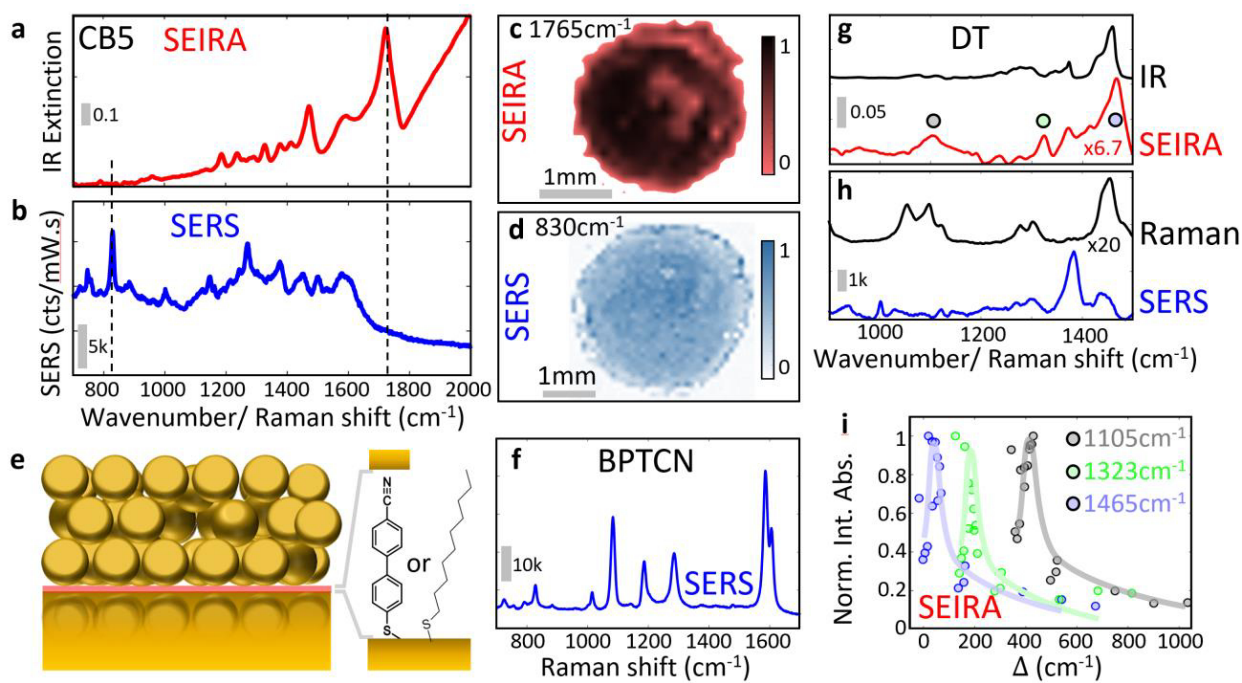

**Figure 4. Comparison of SERS and SEIRA.** (a) FTIR reflectance spectrum (in wavenumbers) and (b) SERS spectrum (in Raman shift, $\lambda_{exc} = 785$ nm), of NP7ML-on-mirror films ($D$=100nm). (c,d) Simultaneous SERS and SEIRA measurements on NP7ML films ($D$=100nm): (c) SEIRA map of CB[5] C=O stretch (1765 cm$^{-1}$), (d) SERS map of CB[5] ring deformation (830 cm$^{-1}$) (normalized peak areas). (e) Schematic of NP3ML-on-mirror film with self-assembled monolayer of 4'-cyanobiphenyl-4-thiol (BPTCN) or decane-1-thiol (DT) in the mirror gap (red), with (f) BPTCN SERS spectrum. (g,h) SEIRA and SERS of monolayer DT in NP7ML-on-mirror films compared to reference liquid IR/Raman spectra[42]. All spectra are background corrected (details in SI Section S10), grey bars give counts/mW.sec for Raman/SERS and absorbance for IR/SEIRA. Labels on SEIRA and Raman spectra indicate intensity scaling. (i) Integrated IR absorbance for DT modes at 1105 cm$^{-1}$ (integrated from 1050-1150 cm$^{-1}$), 1323 cm$^{-1}$ (1300-1340 cm$^{-1}$), and 1465 cm$^{-1}$ (1450-1600 cm$^{-1}$) $vs$ detuning of plasmon mode from each vibration (normalized, lines are guides to the eye).

Sensing analytes with SEIRA is possible by flowing[31] them through the CB[5]-scaffolded nanogaps[43] (preliminary data in SI Section S13, Fig. S25). However besides the vibrational signals from these nanogaps, it is possible to detect molecular self-assembled monolayers (SAMs) placed onto the mirror, before the NP$n$ML is deposited (Fig. 4e, pink molecular layer). Here, we use a robust SAM molecule 4′-cyanobiphenyl-4-thiol (BPT-CN) as a spacer, because of its clearly distinguished vibrational modes from CB[5]. The enhancement of the local fields in the gaps between a NP1ML film and the mirror is clearly seen by the strong SERS from this molecular layer (Fig. 4f), with SERS enhancement factor $1.1 \times 10^6$ (Supplementary Information Section S7). A perfectly ordered AuNP superlattice on a mirror has a field null enforced by boundary conditions on the surface, with poor enhancement of SERS signals expected at the superlattice-mirror gap. In contrast, the NPnML-on-mirror films display strong simultaneous SERS and SEIRA enhancements which arises due to the disordered nature of the AuNP layers allowing for high-angle coupling into the gap between the NP$n$ML layer and the mirror via each defect. Supercell calculations of defects in the ordered superlattice confirm this enhanced coupling (Supplementary Information Section S4 and Fig. S12).

To demonstrate the strong enhancements possible here, we also demonstrate sub-picolitre sensing of 1-decanethiol (DT) SAM (0.07 pL of DT in IR microscope spot size) which is typically hard to detect. The SERS and SEIRA spectra (Fig. 4g,h) of the adsorbed DT are extracted by subtracting the spectrum of the bare film and correcting for the plasmonic background (Supplementary Information Section S10). The SERS and SEIRA enhancement factors for 1-decanethiol are $1.2 \times 10^4$ and $8.3 \times 10^3$ respectively. The observed SEIRA peaks possess different maximum enhancement as a function of relative detuning ($\Delta$) from the plasmon peak ($\omega_{\text{BPP}}$). This emphasizes the role of local plasmonic fields, as different vibrations couple in different ways to the local multipolar plasmonic field, resulting in different near field enhancements.

It is remarkable that the disordered NP$n$ML-on-mirror films display similar bulk plasmon polariton modes to their highly-ordered parent structures, but with much better SERS and SEIRA enhancement factors[19]. Previous theoretical results implied highly ordered structures[44] are needed for dipole-dipole coupling to create the collective plasmonic mode[26]. However, in our case the infrared response is no longer dependent on the precise NP structure, but instead behaves as an effective metamaterial in the homogeneous regime, similar to random AuNP aggregate coupling to ITO plasmons in the near-infrared[45]. Our work shows that structural order is not necessary, and in fact aids sensing applications as it allows more light to be in/out-coupled.

**Conclusions**
In summary, multilayer films of amorphously-arranged AuNPs (NP$n$ML-on-mirror) are fabricated by self-assembly with a precision cucurbit[5]uril molecular glue. The resulting metamaterial displays tunable resonances in the mid-infrared sensing region ($\lambda$=2-10 μm). Our key advance is to show that near- to mid-infrared can be compressed into mode volumes $V_m \sim 10^{-8} \lambda^3$, resulting in strongly enhanced vibrational signatures from gap molecules and plasmonic resonances which are robust to global structural disorder. Instead of using complex microfabricated structures which require combined lithography of µm length-scale antennas with nm gaps, we here utilize facile self-assembly of commercially-available AuNPs with molecular scaffolding to define the optimum architecture with precise sub-nanometre gap sizes.

NP$n$ML-on-mirror films display dual SERS and SEIRA enhancements, enabling sensing of molecules with different spectral features in Raman and IR, and can avoid visible fluorescence backgrounds that contaminate Raman spectra. The strong mid-infrared mode, which arises from a bulk plasmon polariton, allows for both significant SEIRA (EF~$10^6$) and SERS (EF~$10^6$) enhancements, with vibration-plasmon coupling strengths on par with those in vibrational strong coupling. The SERS and SEIRA enhancement factors exceed many state-of-the-art resonant plasmonic antenna structures[3, 30-32] used to confine mid-infrared light for sensing. The strength of the coupling can be attributed to the efficiency with which the NP$n$ML-on-mirror confines MIR light and squeezes it into the locally-ordered nanocavity gaps. The large Purcell factor and extremely small mode volume implies strong confinement of the mid-infrared field within a small number of inter-particle gaps, approaching the limit of confinement possible within such a nanostructure.

The simplicity of this self-assembly enables the creation of robust amorphous metasurfaces. Rather than being a burden, disorder is a strength as it aids diffusion of molecules into the gaps for sensing and permits access to strong optical fields between the NP$n$ML film and the mirror, potentially allowing real time SEIRA measurements within microfluidic flow cells (Fig. S25) with a reusable substrate (Fig. S26). Amorphous NP$n$ML-on-mirror films with extremely small mode volumes open up a key platform with broader uses for infrared radiation capture in increasing the efficiency of IR photodetectors[10], controlling the spectral reflectance of thermal management surfaces[8, 9], in display technologies[46], to modulate IR emission, and vibrational strong coupling-controlled chemistry[47]. The dual enhancements in the visible and IR regions also have the potential to increase nonlinear frequency upconversion[48] or frequency mixing[49] efficiencies.

**METHODS:**

**Multilayer aggregate formation.** Disordered AuNP films were formed by adding 500 µL of gold nanoparticles (20-100 nm, citrate capped, BBI Solutions; 200 nm Sigma-Aldrich) to 500 µL of chloroform in an Eppendorf tube. Aggregation was initiated by the addition of 150 µL of CB[5] (cucurbituril-5-hydrate, Sigma Aldrich), followed by immediate vigorous shaking for 1 minute. After shaking, the solution was left to settle for another minute, causing the separation of the immiscible chloroform and aqueous phases. The aggregated gold nanoparticles then settled to the interface between the two phases. The aqueous phase was thrice-washed by the addition and removal of 500 µL of water to dilute the citrate salts. The aqueous phase was then slowly concentrated by removal of water till a dense Au aggregate bead was formed in chloroform. The bead was then deposited on a gold coated glass substrate (thermally evaporated 5 nm Cr and 500 nm Au) and left to dry. Upon drying, the substrate was cleaned with water, isopropanol, and dried in nitrogen gas s. SEIRA measurements were performed on monolayers of decane-1-thiol (DT, Sigma-Aldrich) prepared by immersion of Au substrates for 30 hours in DT liquid, and subsequent rinsing and cleaning in ethanol and dried in nitrogen gas flow. SERS measurements of 4'-cyanobiphenyl-4-thiol (BPTCN, Sigma-Aldrich) was performed by deposition of AuNP multilayers on a self-assembled monolayer prepared by overnight immersion of gold substrate in 1mM BPTCN in ethanol, and subsequent rinsing in ethanol and drying in nitrogen gas flow.

**Broadband reflectance spectroscopy.** For large scale extinction spectra, a custom-built setup was used to simultaneously measure the visible (Ocean Optics QEPro from 300 nm to 1000 nm), near-infrared (Ocean Optics NIRQuest from 700 nm to 2000 nm), and mid-infrared (FTIR Interspec 402-X from 1700 nm to 20,000 nm) spectrum, using a thermal globar lamp source and ZnSe beam-splitters, and referenced to a clean gold mirror.

**Infrared microscopy.** FTIR reflectance spectral mapping (Shimadzu AIM-9000 FTIR microscope, liquid nitrogen cooled MCT detector, Cassegrain 15X objective 0.7 N.A., 20 μm spot size, Happ-Genzel apodization) was performed, referenced to a clean gold mirror.

**Raman microscopy.** Raman measurements (Renishaw inVia microscope, 20X objective, 785 nm, 0.1 mW incident power) were collected with a 1 second integration time.

**Characterization.** Sample morphology was analysed using scanning electron microscopy (Hitachi S-5500, 10kV accelerating voltage) and white light vertical scanning profilometry (Bruker Contour GTK) to obtain heights of monolayer and bilayer AuNP film regions. Atomic force microscopy (Asylum Research MFP-3D Origin+) was also used to characterize layer heights. X-ray photoelectron spectroscopy (ThermoFisher Escalab 250Xi) was used to measure the surface chemistry of MLAgg films using a monochromated Al Kα X-ray source.

**Simulations.** Finite-difference time-domain (FDTD) simulations (Lumerical FDTD Solutions) of perfectly ordered AuNP superlattices were conducted (for details see SI). Mie scattering simulations were performed with MSTM[50] (multiple-sphere T-matrix formalism) with the Au dielectric function of Johnson & Christy (details in SI).

ASSOCIATED CONTENT

**Supporting Information**.

AUTHOR INFORMATION

**Corresponding Author**

* Prof Jeremy J Baumberg, jjb12@cam.ac.uk

**Contributions**

R.A and J.J.B conceived and designed the experiments. R.A. performed the fabrication and spectroscopic experiments with input from D.B.G., R.C., and A.X. R.A. analysed the data with input from R.C. and and N.S.M.. R.A. and N.S.M. carried out the simulations and the analytical modelling. E.M. performed experiments and analysed the data from IR flow cell measurements with input from R.A. and T.G.E. R.A. and J.J.B. wrote the manuscript with input from all authors.

**Notes**

The authors declare no competing financial interest.

ACKNOWLEDGMENT

We acknowledge support from European Research Council (ERC) under Horizon 2020 research and innovation programme THOR (Grant Agreement No. 829067), and PICOFORCE (Grant Agreement No. 883703). R.A. acknowledges support from the Rutherford Foundation of the Royal Society Te Apārangi of New Zealand, and the Winton Programme for the Physics of Sustainability. R.C. and R.A. acknowledge support from Trinity College, University of Cambridge. The authors thank Dr. Isabella Miele for assistance in analysing SEM images and Dr. Carmen Fernandez-Posada for assistance in collecting X-ray photoelectron spectra. N.S.M. acknowledges support from the German National Academy of Sciences Leopoldina. The authors acknowledge use of the XPS System, part of Sir Henry Royce Institute Cambridge Equipment, EPSRC grant EP/P024947/1. J.J.B, E.M. and T.G.E acknowledge support from the Faraday Institution grant no. FIRG001.